\documentclass[11pt]{article}
\usepackage{latexsym,amsmath,amsfonts,graphicx,subfigure}

\begin{document}
\begin{center}
{\LARGE\textbf{Random consensus in nonlinear systems under fixed topology}}\\
\bigskip
\bigskip
Radha F. Gupta\footnote{Department of Mathematics, University of
Mumbai, India. email: \texttt{guptarf@rediffmail.com}}\\

Poom Kumam\footnote{Institute of Mathematics, Thai Academy of Science and Technology, Thailand. email: \texttt{pkumam@yt.ac.th}}\\

\end{center}

\begin{abstract}

This paper investigates the consensus problem in almost sure sense
for uncertain multi-agent systems with noises and fixed topology. By
combining the tools of stochastic analysis, algebraic graph theory,
and matrix theory, we analyze the convergence of a class of
distributed stochastic type non-linear protocols. Numerical examples
are given to illustrate the results.

\bigskip

\end{abstract}

\bigskip
\normalsize
\newpage

\noindent{\Large\textbf{1. Introduction}}
\smallskip

In the distributed control problems, a critical problem is to design
distributed protocols such that group of agents can achieve
consensus via local communications. Distributed coordination for
multi-agent systems has become a very active research topic and
attracted great attention of researchers in recent years; see e.g.
\cite{13,1,6,2,3,5}. The network protocol is an interaction rule,
which ensures the whole group can achieve consensus on the shared
data in a distributed manner. Consensus problems cover a very broad
spectrum of applications including formation control, distributed
computation, unmanned aerial vehicles, mobile robots, autonomous
underwater vehicles, distributed filtering, multi-sensor data
fusion, automated highway systems, and formation control of
satellite clusters \cite{4}.

A key problem of the consensus problem is the convergence time. Some
researchers have investigated the so-called finite time consensus,
where the consensus occurs in a finite time, see e.g.
\cite{10,17,2,7,8,40,41,16,9,50,51,52}. However, in most of these
works, the random noises have not been considered. Since the random
noises are inevitable in the nature, we must take them into account.

In this paper, we investigate a continuous-time nonlinear
multi-agent system with random noises. We provide extensive
simulation results to show the finite-time consensus as well as the
effect of the randomness.

The rest of the paper is organized as follows. In Section 2, we
provide some preliminaries and formulate the problem. Section 3
contains the numerical simulations and we draw conclusion in Section
4.

\bigskip
\noindent{\Large\textbf{2. Problem formulation}}
\smallskip

Let
$\mathcal{G}(\mathcal{A})=(\mathcal{V}(\mathcal{G}),\mathcal{E}(\mathcal{G}),\mathcal{A})$
be a weighted directed graph with the set of vertices
$\mathcal{V}(\mathcal{G})=\{1,2,\cdots,n\}$ and the set of arcs
$\mathcal{E}(\mathcal{G})\subseteq\mathcal{V}(\mathcal{G})\times\mathcal{V}(\mathcal{G})$.
The vertex $i$ in $\mathcal{G}(\mathcal{A})$ represents the $i$th
agent, and a directed edge $(i,j)\in\mathcal{E}(\mathcal{G})$ means
that agent $j$ can directly receive information from agent $i$, the
parent vertex. The set of neighbors of vertex $i$ is denoted by
$\mathcal{N}(\mathcal{G},i)=\{j\in\mathcal{V}(\mathcal{G})|\
(j,i)\in\mathcal{E}(\mathcal{G})\}$. The corresponding graph
Laplacian $L(\mathcal{A})=(l_{ij})\in\mathbb{R}^{n\times n}$ can be
defined as
$$
l_{ij}=\left\{\begin{array}{ll}\sum_{k=1,k\not=n}^{n}a_{ik},&j=i\\
-a_{ij},&j\not=i
\end{array}\right..
$$
If $\mathcal{A}^T=\mathcal{A}$, we say $\mathcal{G}(\mathcal{A})$ is
undirected.

We study a system consisting of $n$ dynamic agents, indexed by
$1,2,\cdots,n$. The interaction topology among them are described by
the weighted directed graph $\mathcal{G}(\mathcal{A})$ as defined
above. We further assume the diagonal entries of $\mathcal{A}$ are
zeroes. The continuous-time dynamics of $n$ agents is described as
follows:
\begin{equation}
\dot{x}_i(t)=u_i(t),\quad i=1,2,\cdots,n,\label{1}
\end{equation}
where $x_i(t)\in\mathbb{R}$ is the state of the $i$th agent, and
$u_i(t)\in\mathbb{R}$ is the control input. Denote
$x(t)=(x_1(t),\cdots,x_n(t))^T$ and $1=(1,\cdots,1)^T$ with
compatible dimensions. For a vector $z\in\mathbb{R}^n$, let
$\|z\|_{\infty}$ denote its $l^{\infty}$-norm.

Given a protocol $\{u_i : i=1,2,\cdots,n\}$, the multi-agent system
is said to solve a consensus problem if for any initial states and
any $i,j\in\{1,\cdots,n\}$, $|x_i(t)-x_j(t)|\rightarrow0$ as
$t\rightarrow\infty$; and it is said to solve a finite-time
consensus problem if for any initial states, there is some
finite-time $t^*$ such that $x_i(t)=x_j(t)$ for any
$i,j\in\{1,\cdots,n\}$ and $t\ge t^*$.

We consider the following protocol:
\begin{equation}
u_i=f_i\bigg(\sum_{j\in\mathcal{N}(\mathcal{G}(\mathcal{A}),i)}a_{ij}(x_j-x_i)\bigg),\label{2}
\end{equation}
where functions $f_i:\mathbb{R}\rightarrow\mathbb{R}$,
$i=1,\cdots,n$ is taken as a random bounded continuous function.

\bigskip
\noindent{\Large\textbf{3. Main result}}
\smallskip

We first provide a key result from \cite{2}.

\smallskip
\noindent\textbf{Theorem 1.} \itshape \quad Assume
$\mathcal{G}(\mathcal{A})$ is a
directed graph with Laplacian matrix $L(\mathcal{A})$, then we have\\
(i)\quad $L(\mathcal{A})1=0$ and all non-zero eigenvalues have
positive real
parts;\\
(ii)\quad $L(\mathcal{A})$ has exactly one zero eigenvalue if and
only if $\mathcal{G}(\mathcal{A})$ has a spanning tree;\\
(iii)\quad If $\mathcal{G}(\mathcal{A})$ is strongly connected, then
there is a positive column vector $\omega\in\mathbb{R}^n$ such that
$\omega^TL(\mathcal{A})=0$;\\
(iv)\quad Let $b=(b_1,\cdots,b_n)^T$ be a nonnegative vector and
$b\not=0$. If $\mathcal{G}(\mathcal{A})$ is undirected and
connected, then $L(\mathcal{A})+diag(b)$ is positive definite. Here,
$diag(b)$ is the diagonal matrix with the $(i,i)$ entry being
$b_i$.\normalfont
\smallskip

The state trajectories of the agents are shown in Fig. 1 to Fig. 5.
In Fig. 1, we take $f_i(x)=i*sgn(x)$. In Fig. 2, we take
$f_i(x)=\sqrt{i}*sgn(x)$. In Fig. 3, we take $f_i(x)=i^2*sgn(x)$. In
Fig. 4, we take $f_i(x)=\sin(i)*sgn(x)$. In Fig. 5, we take
$f_i(x)=\cos(i)*sgn(x)$.

\begin{figure}[htb]
\centering {\includegraphics[width=.7\textwidth]{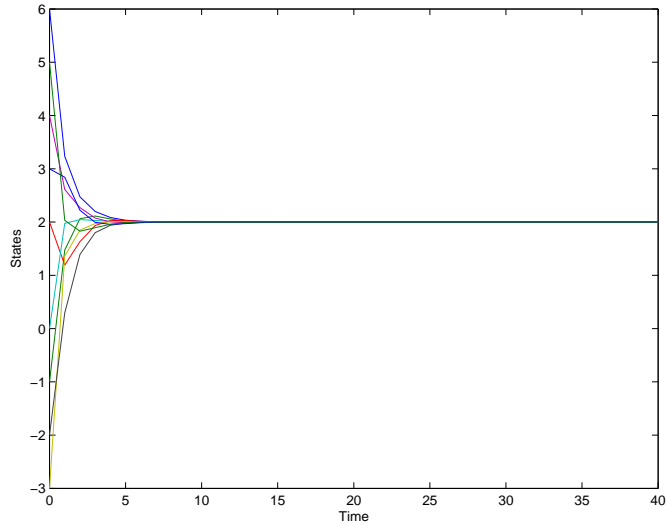}}
\caption{The state trajectories of the agents.}
\end{figure}

\begin{figure}[htb]
\centering {\includegraphics[width=.7\textwidth]{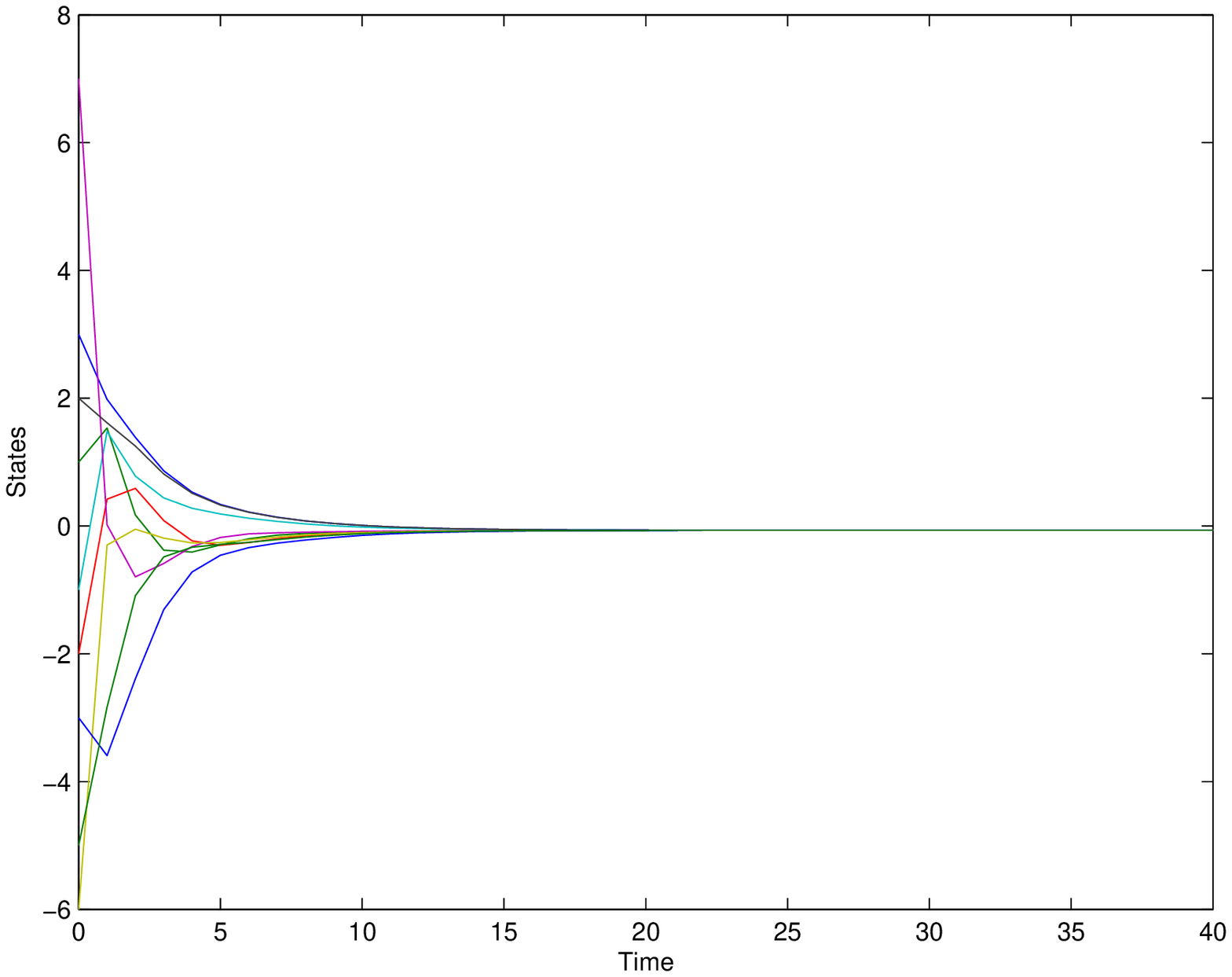}}
\caption{The state trajectories of the agents.}
\end{figure}

\begin{figure}[htb]
\centering {\includegraphics[width=.7\textwidth]{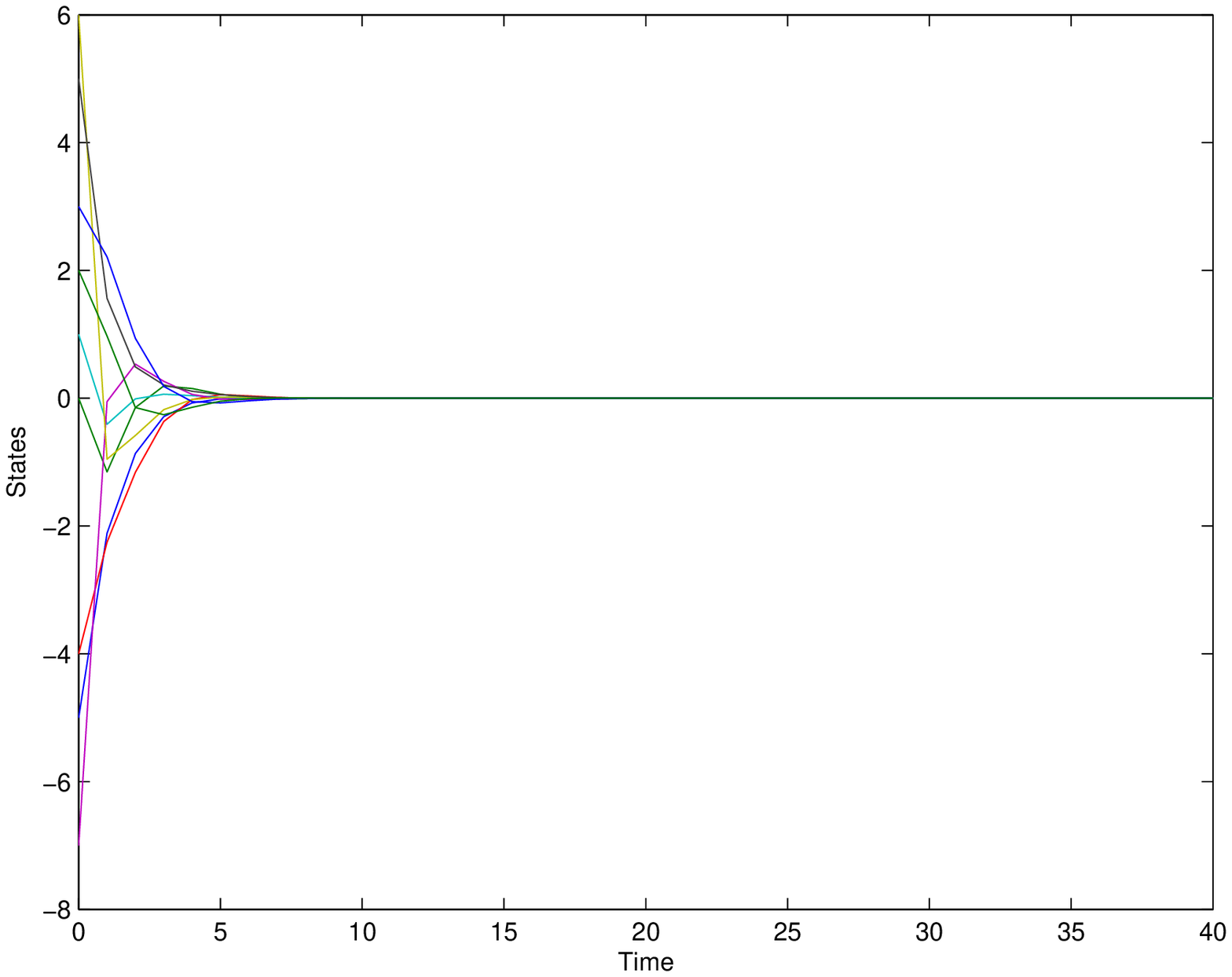}}
\caption{The state trajectories of the agents.}
\end{figure}

\begin{figure}[htb]
\centering {\includegraphics[width=.7\textwidth]{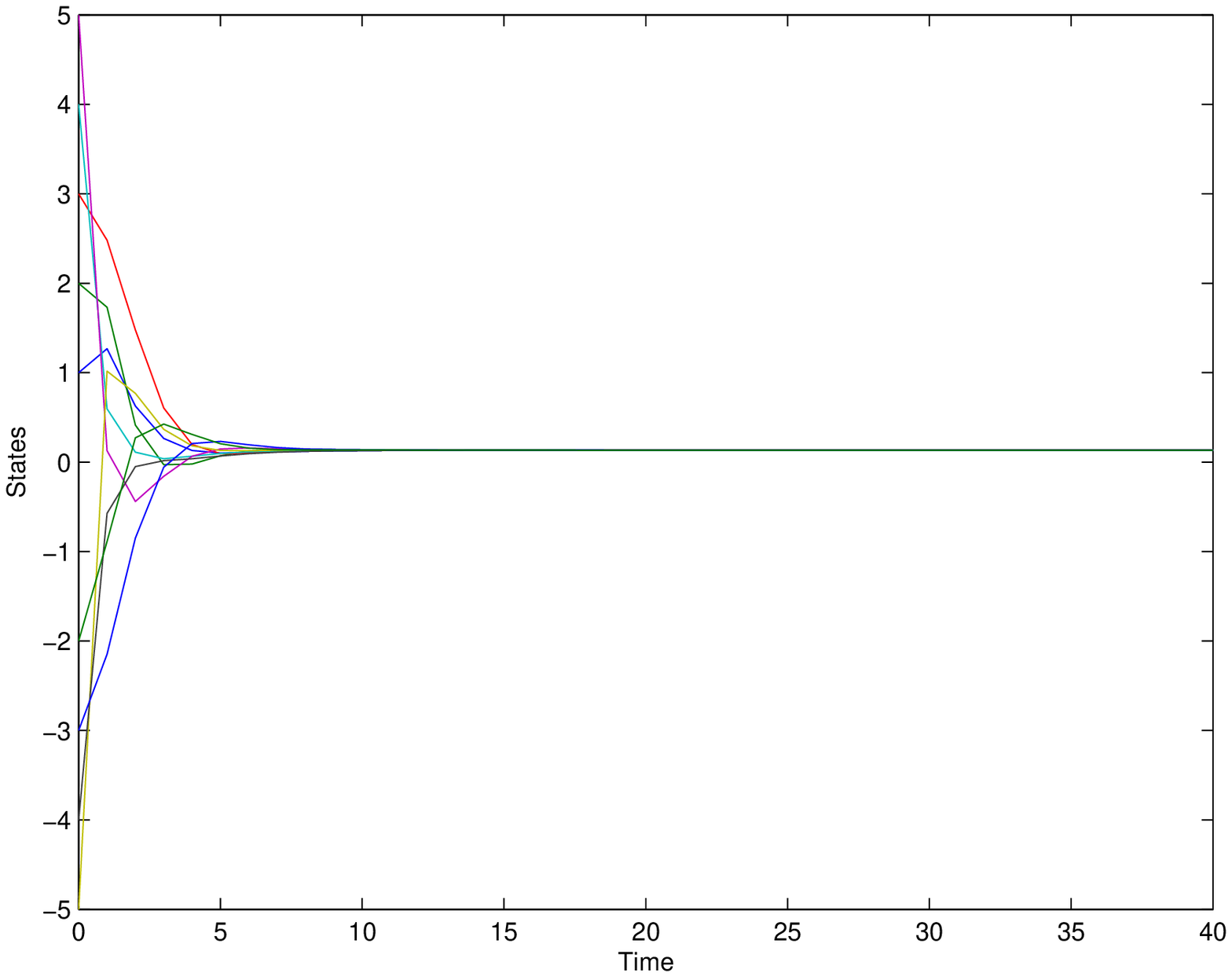}}
\caption{The state trajectories of the agents.}
\end{figure}

\begin{figure}[htb]
\centering {\includegraphics[width=.7\textwidth]{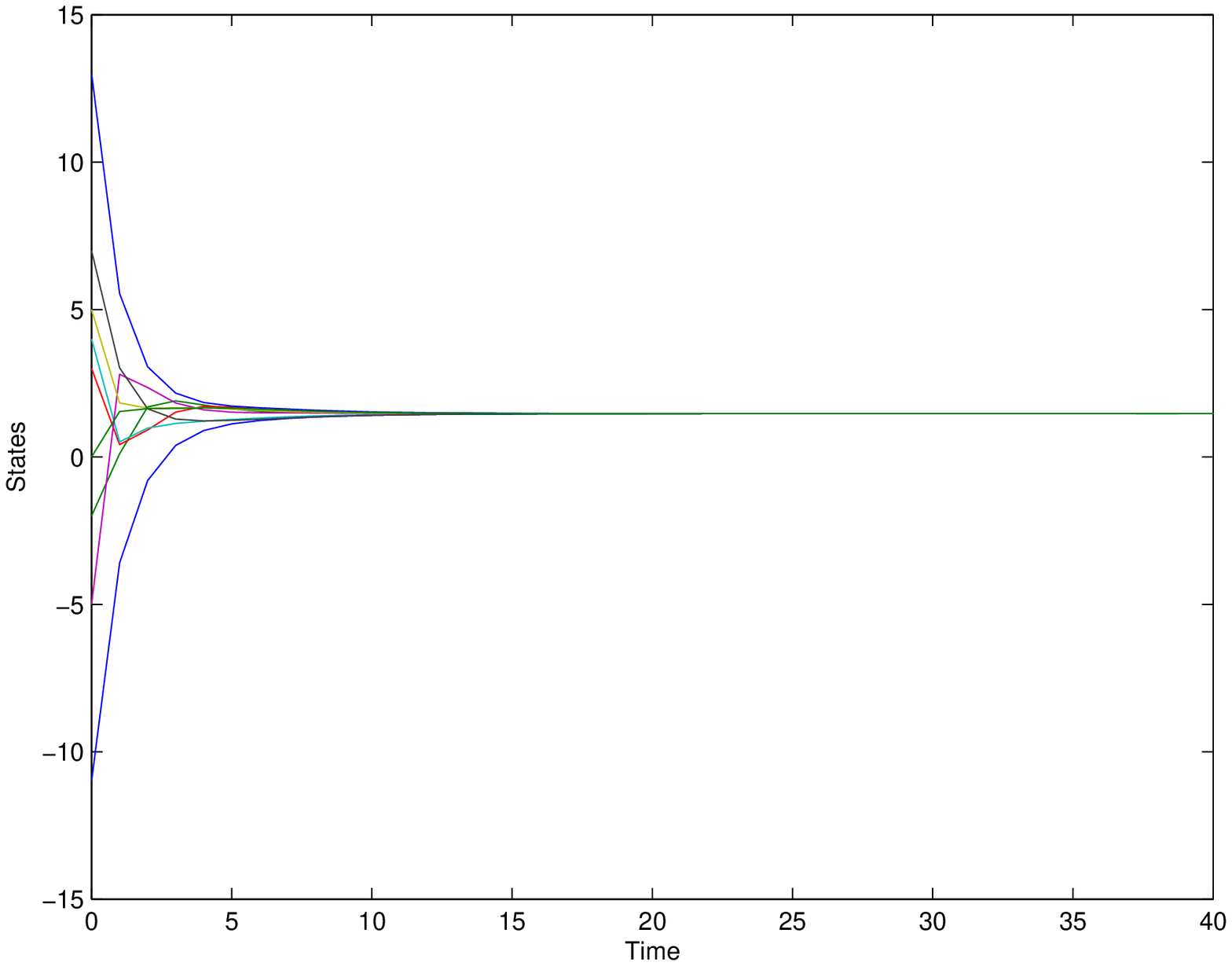}}
\caption{The state trajectories of the agents.}
\end{figure}

\bigskip
\noindent{\Large\textbf{4. Conclusions}}
\smallskip

This paper attempted to look for some insight into the behavior of
random consensus problem on a fixed network. We consider various
random networks and protocols. The simulations show that the systems
achieve finite consensus quite fast despite of random noises. For
future researches, we will focus on the switching topologies as well
as coupling time delays.


\bigskip
\begin{thebibliography}{10}
\bibitem{14}S.\ P.\ Bhat, D.\ S.\ Bernstein, Finite-time stability
of continuous autonomous systems. \textit{SIAM J. Control Optim.},
38(2000) 751-766
\bibitem{10}J.\ Cort\'es, Finite-time convergent gradient flows with
applications to network consensus. \textit{Automatica}, 42(2006)
1993-2000
\bibitem{17}J.\ Cort\'es, Distributed algorithms for reaching consensus on general
functions. \textit{Automatica}, 44(2008) 726-737
\bibitem{15}V.\ T.\ Haimo, Finite time controllers. \textit{SIAM J. Control
Optim.}, 24(1986) 760-770
\bibitem{52}H.\ Hao, P.\ Barooah, Improving convergence rate of distributed consensus through
asymmetric weights. \textit{arXiv:1203.2717}, 2012
\bibitem{13}Y.\ Hatano, M.\ Mesbahi, Agreement over random networks.
\textit{IEEE Trans. Autom. Control}, 50(2005) 1867-1872
\bibitem{16}Q.\ Hui, W.\ M.\ Haddad, S.\ P.\ Bhat, Semistability theory for differential inclusions with applications to consensus problems in dynamical networks with switching
topology. \textit{Proc. American Control Conference}, 2008 3981-3986
\bibitem{9}Q.\ Hui, W.\ M.\ Haddad, S.\ P.\ Bhat, Finite-time
semistability and consensus for nonlinear dynamical networks.
\textit{IEEE Trans. on Autom. Control}, 53(2008) 1887-1900
\bibitem{1}A.\ Jadbabaie, J.\ Lin, A.\ S.\ Morse, Coordination of
groups of mobile autonomous agents using nearest neighbor rules.
\textit{IEEE Trans. Autom. Control}, 48(2003) 988-1001
\bibitem{6}R.\ Olfati-Saber, Flocking for multi-agent dynamic
systems: Algorithms and theory. \textit{IEEE Trans. on Autom.
Control}, 51(2006) 401-420
\bibitem{4}R.\ Olfati-Saber, J.\ A.\ Fax, R.\ M.\ Murray, Consensus
and cooperation in networked multi-agent systems.
\textit{Proceedings of the IEEE}, 95(2007) 215-233
\bibitem{2}R.\ Olfati-Saber, R.\ M.\ Murray, Consensus problems in
networks of agents with switching topology and time-delays.
\textit{IEEE Trans. Autom. Control}, 49(2004) 1520-1533
\bibitem{3}W.\ Ren, R.\ W.\ Beard, Consensus seeking in multi-agent
systems under dynamically changing interaction topologies.
\textit{IEEE Trans. on Autom. Control}, 50(2005) 655-661
\bibitem{5}W.\ Ren, R.\ W.\ Beard, E.\ M.\ Atkins, Information
consensus in multivehicle cooperative control. \textit{IEEE Control
Systems Magazine}, 27(2007) 71-82
\bibitem{51}Y.\ Shang, Finite-time consensus for multi-agent systems with
fixed topologies. \textit{Internat. J. Systems Sci.}, 43(2012)
499-506
\bibitem{50}J.\ Shen, J.\ Cao, Finite-time synchronization of coupled neural
networks via discontinuous controllers. \textit{Cognitive
Neurodynamics}, 5(2011) 373-385
\bibitem{40}L.\ Wang, Z.\ Chen, Z. Liu, Z.\ Yuan, Finite time agreement
protocol design of multi-agent systems with communication delays.
\textit{Asian Journal of Control}, 11(2009) 281-286
\bibitem{41}Z.\ Wang, S.\ Li, S.\ Fei, Finite-time tracking control of a nonholonomic mobile
robot. \textit{Asian Journal of Control}, 11(2009) 344-357
\bibitem{8}L.\ Wang, F.\ Xiao, Finite-time consensus problems for
networks of dynamic agents. \textit{IEEE Transactions on Automatic
Control}, 55(2010) 950-955
\bibitem{7}F.\ Xiao, L.\ Wang, Y.\ Jia, Fast information sharing in
networks of autonomous agents. \textit{Proc. American Control
Conference}, 2008 4388-4393


\end{thebibliography}
\end{document}